\documentclass[amsmath,amssymb,pre,twocolumn,superscriptaddress]{revtex4}

\usepackage{graphicx}
\usepackage{bm}
\usepackage{color,graphics}
\usepackage{soul}

\begin{document}

\title{Collisionless shock formation, spontaneous electromagnetic fluctuations\\and streaming instabilities}

\author{A. Bret}
\affiliation{ETSI Industriales, Universidad de Castilla-La Mancha, 13071 Ciudad Real, Spain}
 \affiliation{Instituto de Investigaciones Energ\'{e}ticas y Aplicaciones Industriales, Campus Universitario de Ciudad Real,  13071 Ciudad Real, Spain.}

 \author{A. Stockem}
\affiliation{GoLP/Instituto de Plasmas e Fus\~{a}o Nuclear - Laborat\'{o}rio Associado,
Instituto Superior T\'{e}cnico, Lisboa, Portugal}

 \author{F. Fiuza}
\affiliation{GoLP/Instituto de Plasmas e Fus\~{a}o Nuclear - Laborat\'{o}rio Associado,
Instituto Superior T\'{e}cnico, Lisboa, Portugal}

 \author{C. Ruyer}
\affiliation{CEA, DAM, DIF F-91297 Arpajon, France}

 \author{L. Gremillet}
\affiliation{CEA, DAM, DIF F-91297 Arpajon, France}

 \author{R. Narayan}
\affiliation{Harvard-Smithsonian Center for Astrophysics,
60 Garden Street, MS-51 Cambridge, MA 02138, USA}

 \author{L.O. Silva}
\affiliation{GoLP/Instituto de Plasmas e Fus\~{a}o Nuclear - Laborat\'{o}rio Associado,
Instituto Superior T\'{e}cnico, Lisboa, Portugal}

\begin{abstract}
Collisionless shocks are ubiquitous in astrophysics and in the lab. Recent numerical simulations and experiments have shown how they can arise from the encounter of two collisionless plasma shells. When the shells interpenetrate, the overlapping region turns unstable, triggering the shock formation. As a first step towards a microscopic understanding of the process, we analyze here in detail the initial instability phase.  On the one hand, 2D relativistic PIC simulations are performed where two symmetric initially cold pair plasmas collide. On the other hand, the instabilities at work are analyzed, as well as the field at saturation and the seed field which gets amplified. For mildly relativistic motions and onward, Weibel  modes govern the linear phase. We derive an expression for the duration of the linear phase in good agreement with the simulations. This saturation time constitutes indeed a lower-bound for the shock formation time.
\end{abstract}


\maketitle

\section{Introduction}
Colliding plasma shells are present in a variety of physical settings. Astrophysical jets produced by black holes are expected to generate a shock when interacting with the interstellar medium \cite{Begelman1984,Harris2006}. Still in astrophysics, the Fireball scenario for Gamma-Rays-Bursts \cite{Piran2004,Nakar2007} relies on shock particle acceleration \cite{Bell1978a,Bell1978b,Blandford78}, where the shock arises from the encounter of two ultra-relativistic plasma blobs ejected from a central engine.  Non-relativistic Supernova Remnant Shocks are also instrumental in accelerating high energy cosmic rays \cite{NatureCRSnR,Dieckmann2000}.

For the collisionless environments considered, Particle-In-Cell (PIC) simulations are an efficient tool to study these processes. The formation of a shock following the collision of two plasmas was first explored in Ref. \cite{SilvaApJ} and observed in Ref. \cite{Spitkovsky2005}. Subsequent particle acceleration has been observed in numerous simulations  \cite{Hededal2004,Nishikawa2005,chang2008,Spitkovsky2008,Spitkovsky2008a,Martins2009}. In addition, shock generation through counter-streaming plasmas has already been observed in laboratory \cite{Kuramitsu2011,Joseph2011,CollLess_NJP_2011,ross2012,Gregori2012}, and the conditions required to drive near relativistic shocks have been identified \cite{Fiuza2012}. The corresponding particle acceleration could be a promising alternative to current plasma accelerator schemes \cite{NatureGolp}.

\begin{figure}
\begin{center}
\includegraphics[width=0.45\textwidth]{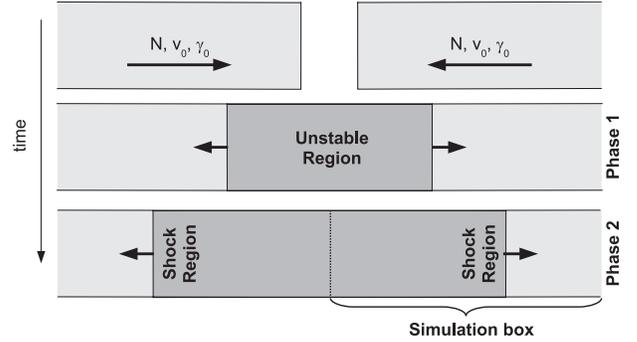}
\end{center}
\caption{Phases of the shock formation. Two identical pair plasmas interpenetrate. The overlapping region turns unstable, and two shocks form near the border of each shell. The simulation box contains half of the system.} \label{fig:setup}
\end{figure}

Although the full shock formation process has thus been now repeatedly observed, a first principle understanding of the very birth of the shock is still lacking. Such an understanding could provide an accurate timing of the shock formation time, and constraints the conditions required to form a shock in the first place. Whether they are in the lab, in a computer or in the vicinity of a supernova, it should be possible to separate the scenario leading to the shock into two phases represented schematically on Figure \ref{fig:setup}. In the first phase, plasma shells  make contact, then overlap, and the overlapping region turns unstable. An instability grows and saturates. At this junction, the total density in the overlapping region is roughly the sum of each plasma density. A second phase is therefore needed during which nonlinear processes pick-up the system from the end of the linear phase, and build-up the Rankine-Hugoniot expected density jump near the borders of the interpenetrating shells.

\begin{figure*}
\begin{center}
\includegraphics[width=0.95\textwidth]{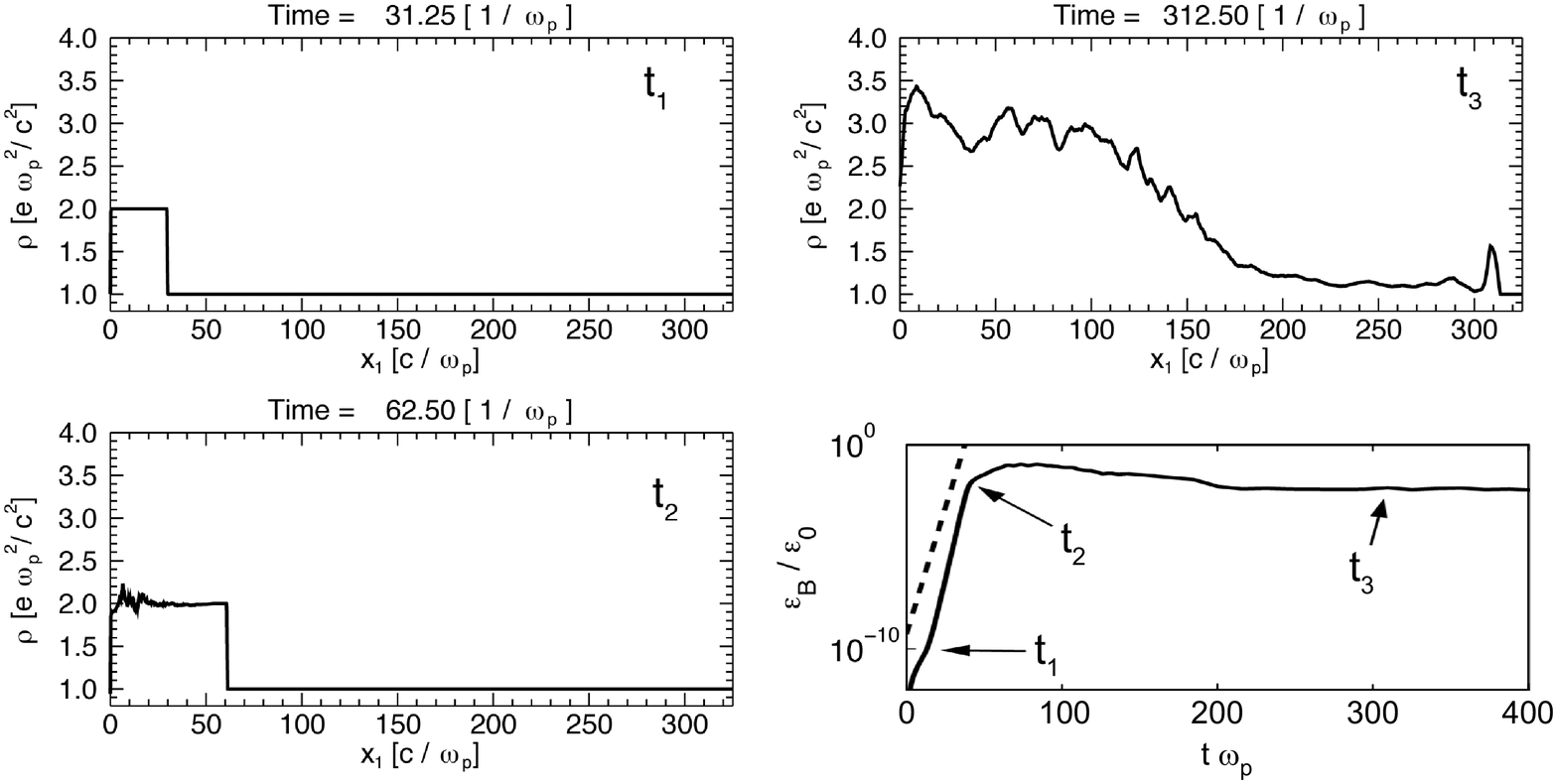}
\end{center}
\caption{Integrated density in the direction normal to the flow for 3 instants of a typical shock formation simulation. The last plot shows the growth of the magnetic energy integrated over the transverse direction, and $x_1\in [0,7\sqrt{\gamma_0}c/\omega_p]$. The dashed line is the theoretical growth-rate. The initial Lorentz factor was $\gamma_0=25$. All the field growth plots look qualitatively the same until $\gamma_0=10^4$. The saturation time $\tau_s$ is $t_2$. The field at saturation is $B(\tau_s)$.} \label{fig:form}
\end{figure*}

\begin{figure*}
\begin{center}
\includegraphics[width=0.95\textwidth]{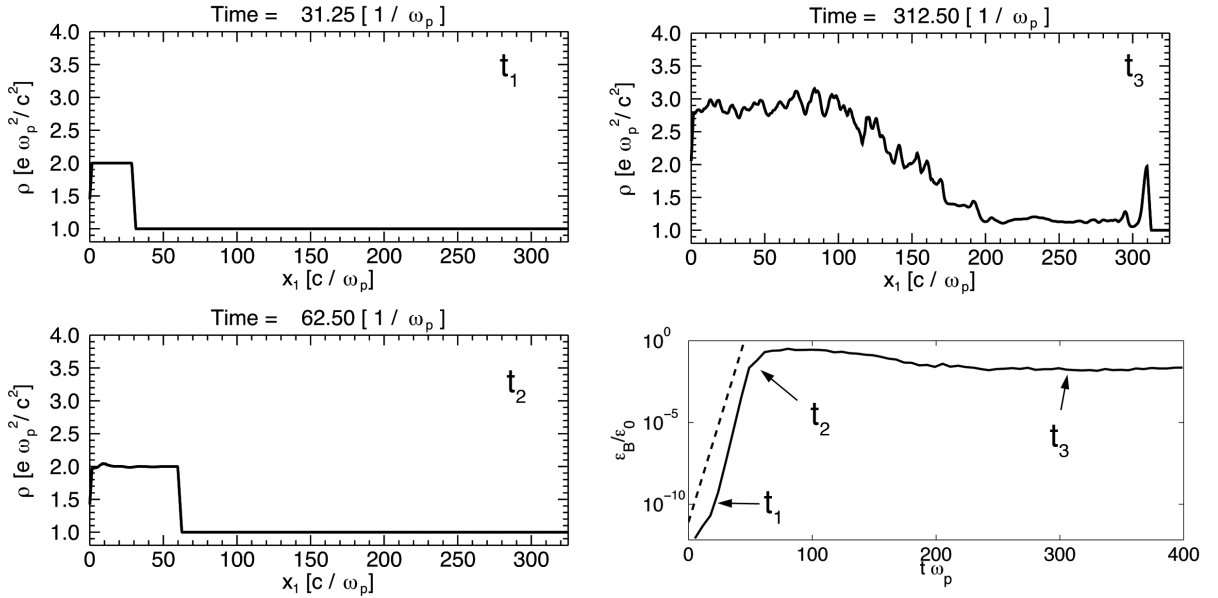}
\end{center}
\caption{Same as Fig. \ref{fig:form}, but using 800 particles per cell.} \label{fig:form800}
\end{figure*}

The present paper is concerned with the first of these two phases. The collision of two identical cold relativistic pair plasmas has been simulated in 2D with the PIC Code \textsc{osiris} \cite{Osiris1,Osiris2}. The details of the simulations are given in Section \ref{sec:sim}. This setup has been chosen for its simplicity, allowing for a direct comparison with theory as the only free parameter is the initial Lorentz factor of the shells $\gamma_0$. In the simulation, a neutral e$^-$/e$^+$ plasma is made to bounce back against a wall and to interact with itself (Fig. \ref{fig:setup}), which enables to describe only half of the symmetric physical system. Periodic boundary conditions are applied in the transverse direction. A series of snapshots from a simulation with $\gamma_0=25$ is displayed on Fig. \ref{fig:form}. Only the right part of the system pictured on Fig. \ref{fig:setup} is showed. These successive plots of the integrated density along the direction normal to the flow clearly show how the overlapping region at twice the upstream density turns unstable, before the shock density jump builds up. We observe in the simulations that in phase 1 the fields grow in a well defined spatial region that extends up to $\sim 7 \sqrt{\gamma_0} c/\omega_p$ from the wall. The saturation time $\tau_s$ ($t_2$ on Figure \ref{fig:form}) is defined as the end of the exponential growth of the field energy integrated over the region $x_1\in [0,7\sqrt{\gamma_0}c/\omega_p]$, where the non-relativistic plasma frequency reads $\omega_p^2 = 4 \pi n_e e^2/ m_e$. The field at saturation is simply $B(\tau_s)$.

As discussed in Sec. \ref{sec:sim}, simulations have been run with 8 particles per cell. Figure \ref{fig:form800} displays the same data as Fig. \ref{fig:form}, but running the simulations with 800 particles per cell. No noticeable qualitative difference appears with respect to our runs.

As will be checked, the instabilities at play can be interpreted in terms of the homogeneous theory for such, although the geometry is finite here, since our shells have one contact open boundary. As it amplifies a seed field from its initial fluctuation value to saturation, the instability governs this first phase of the shock formation process for a time $\tau_s$ that we labeled ``saturation time''. A theoretical determination of this time, in good agreement with the simulations, is the main result of this paper. Not only does $\tau_s$ give a lower bound for the shock formation time, it also sheds a light on the amplitude of the amplified initial fluctuations.

\section{Instability analysis}\label{sec:insta}
\begin{figure}
\begin{center}
\includegraphics[width=0.45\textwidth]{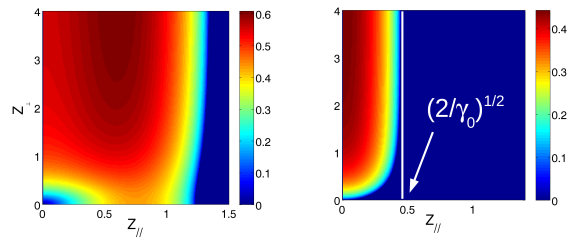}
\end{center}
\caption{(Color Online) Growth-rate in $\omega_p$ units, in terms of $\mathbf{Z}=\mathbf{k} v_0/\omega_p$ for $\gamma_0=1.1$ (\textit{left}) and $\gamma_0=10$ (\textit{right}).} \label{fig:taux}
\end{figure}

Here we deal with the first phase of the shock formation, namely the instability of the overlapping region, where we focus on relativistic shocks. Indeed, if counter streaming collisionless plasmas were not unstable, they would simply go through each other. Let us start ignoring the finite geometry at stake here, and analyze the system as if it were homogeneous. The full unstable $\mathbf{k}$ spectrum has been analyzed long ago in the cold regime, where a shell is much denser than the other \cite{fainberg,Watson,bludman}. These early results were recently generalized to the hot symmetric case \cite{BretPoPHierarchie,BretPoPReview}. For wave-vectors aligned with the flow, we find two-stream unstable modes. For wave-vectors normal to the flow, we find the current filamentation, or Weibel, instability. Finally, modes propagating at arbitrary angle with the flow are also unstable. As the two plasmas penetrate each other, all the modes are excited. But the fastest growing one quickly overcomes the others, and shapes the linear phase. For the case we consider, a calculation of the growth-rate for every possible wave number is pictured on Fig. \ref{fig:taux} for two Lorentz factors $\gamma_0=1.1$ and 10, in terms of the reduced wave-vector,
\begin{equation}
\mathbf{Z}=\frac{\mathbf{k} v_0}{\omega_p},
\end{equation}
where $v_0$ is the initial velocity of the plasmas, and $\omega_p$ the electronic plasma frequency of one of them. The calculation, like the simulation, is conducted in the center of mass reference frame, where the two plasmas come from opposite directions at the same speed. For $\gamma_0=1.1$, the dominant mode is oblique while for $\gamma_0=10$, current filamentation dominates. An in-depth study of the problem found indeed that only these two types of modes can dominate \cite{BretPoPHierarchie,BretPoPReview}. The transition from oblique to filamentation occurs for $\gamma_0=\sqrt{3/2}$, as explained in Appendix \ref{ap:1}. Note that although the analysis of Refs. \cite{BretPoPHierarchie,BretPoPReview} was conducted for counter streaming \emph{electron} beams, counter streaming \emph{pair} beams are linearly equivalent because the linear regime scales like the square of the charge.

We thus find that unless $\gamma_0<\sqrt{3/2}$, current filamentation should govern the interaction with a growth rate,
\begin{equation}\label{eq:gr}
\frac{\delta}{\omega_p}=\frac{v_0}{c}\sqrt{\frac{2}{\gamma_0}}\sim\sqrt{\frac{2}{\gamma_0}}.
\end{equation}
Comparing this value to the growth of the field observed in the overlapping region results in a very satisfactory agreement, as evidenced on Fig. \ref{fig:form}. We have also checked that the Weibel/oblique transition does occur around $\gamma_0=\sqrt{3/2}$. Note that a shock \textit{also} forms for $\gamma_0<\sqrt{3/2}$ (not shown).

This oblique/filamentation transition may seem at odds with previous works on instability hierarchy \cite{BretPoPHierarchie,bretApJ2009,BretPoPReview}, suggesting filamentation always governs the spectrum for symmetric systems. Electrostatic instabilities with parallel wave
vectors have equally been found slower than filamentation for relativistic flows  \cite{Michno2010,Shaisultanov2012}. However, the relevant hierarchy maps, like Fig. 5 of Ref. \cite{bretApJ2009} for example, already showed filamentation does not govern symmetric systems all the way down to $\gamma_0=1$.
Instead, a very little gap was found for oblique electrostatic modes to dominate, between $\gamma_0=1$ and a unspecified value of $\gamma_0>1$. Until now, this little gap did not attract much interest, and it is still overall fair to say that in the relativistic regime, filamentation is the important instability for symmetric systems.

How can a theory developed for an homogeneous system, apply  to the present inhomogeneous system? The instability time scale varies like $\delta^{-1}\propto \sqrt{\gamma_0}/\omega_p$. By the time $\sqrt{\gamma_0}/\omega_p$ after contact, the overlapping region is already $d \sim \sqrt{\gamma_0}c/\omega_p$ wide. But the parallel scale length relative to instabilities is precisely  $\lambda_i=\sqrt{\gamma_0}c/\omega_p$. Even if at the very beginning of the instability process, $d\gg\lambda_i$ is not fulfilled, the strong inequality is quickly realized with time passing, so that most of the instability process develops in a setting fulfilling the homogeneous approximation.

Knowing the growth-rate (\ref{eq:gr}) should allow for an accurate timing of the linear phase. Assuming the instability amplifies a seed field of amplitude $B_i$ up to a saturation level $B_s$, we can write for the saturation time $\tau_s$,
\begin{equation}\label{eq:Bf}
B_f=B_i e^{\delta\tau_s}~~\Rightarrow~~\tau_s=\frac{1}{2\delta}\ln\left(\frac{B_f^2}{B_i^2} \right),
\end{equation}
where, for convenience, we consider the field energy $B^2$ ratio, instead of the field itself. Determining the saturation time amounts then to determine the initial and final fields. We will first discuss the saturation field.

\section{Field at saturation }\label{sec:sat}
One way to derive the value of the saturation field $B_f=B(\tau_s)$, consists in stating that the field grows exponentially as long as it is small enough for the system to fit the linear approximation. Since a field $B_f$ affects particles on a time scale given by the cyclotron frequency, this implies \cite{califano2,Medvedev1999},
\begin{equation}\label{eq:Bs1}
\frac{qB_{f1}}{\gamma_0 m c}=\delta~~\Rightarrow~~B_{f1}= \frac{\gamma_0 m}{q} \delta c.
\end{equation}
Another way of evaluating the field at saturation is to write that as it grows, particles start oscillating in the field of the fastest growing $k_m$ at frequency \cite{davidsonPIC1972,Achterberg2007},
\begin{equation}\label{eq:Bs1-1}
\omega_{B_k}^2=\frac{q v_0 k B}{\gamma_0 mc}\sim \frac{qk_m B}{\gamma_0 m}.
\end{equation}
Here again, linear theory yielding to an exponential growth breaks down when $\omega_{B_k}\sim\delta$, which gives a second value for the field at saturation,
\begin{equation}\label{eq:Bs2}
B_{f2} =  \frac{\gamma_0 m}{q} \frac{\delta^2}{k_m}.
\end{equation}
Finally, one can write the linear approximation breaks down when the Larmor radius of the particles in the growing field equates $k_m^{-1}$. This third criterion thus gives,
\begin{equation}\label{eq:Bs3}
B_{f3}=\frac{\gamma_0 m}{q}c^2 k_m,
\end{equation}
where $v_o\sim c$ has been used. The quantity $k_m$ has been numerically measured with,
\begin{equation}\label{eq:km}
k_m\sim\frac{\omega_p}{c\sqrt{\gamma_0}}.
\end{equation}
Accounting in addition for the growth-rate expression (\ref{eq:gr}) gives,
\begin{eqnarray}
B_{f1}^2&=&2\gamma_0 b^2,\label{eq:fields_f1}\\
B_{f2}^2&=&4\gamma_0 b^2,\label{eq:fields_f2}\\
B_{f3}^2&=&\gamma_0 b^2,\label{eq:fields_f3}
\end{eqnarray}
where,
\begin{equation}
b=\frac{mc\omega_p}{q},
\end{equation}
is the magnetic field unit of the simulations.

\begin{figure}
\begin{center}
\includegraphics[width=0.45\textwidth]{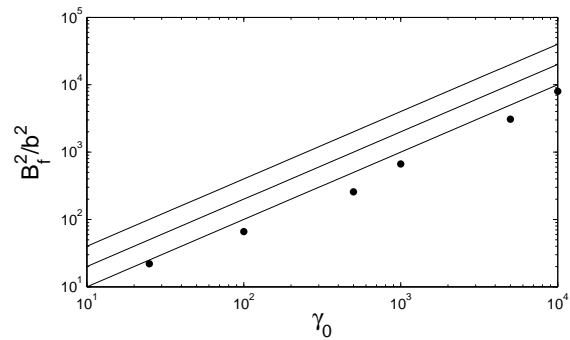}
\end{center}
\caption{Field at saturation from the simulations (\textit{circles}), compared with Eqs. (\ref{eq:fields_f1}-\ref{eq:fields_f3}).} \label{fig:compaBf}
\end{figure}

The magnetic field for the filamentation instability grows like $\sin(ky)e^{\delta t}$. As a result, particles in the vicinity of $y=0$ $[\pi]$ are the ones involved in the second saturation mechanism, described by Eq. (\ref{eq:Bs2}). Particles near $y=\pi$ $[\pi]$ experience the kind of trapping involved with Eqs. (\ref{eq:Bs1},\ref{eq:Bs3}). The linear hypothesis is first broken when the field reaches $\min(B_{f1},B_{f2},B_{f3})=B_{f3}$. Figure \ref{fig:compaBf} compares the field observed in the simulation at the end of the linear phase with Eqs. (\ref{eq:fields_f1}-\ref{eq:fields_f3}). The agreement with Eq. (\ref{eq:fields_f3}) is good and the correct scaling is recovered.
At any rate, a numerical pre-factor cannot play a major role once inserted into the logarithm of Eq. (\ref{eq:Bf}) for the saturation time.

A consequence of the observed $\gamma_0$ scaling is that the field energy relative to the beam one reads,
\begin{equation}
\frac{B_f^2/8\pi}{\gamma_0nmc^2}\sim 1,
\end{equation}
displaying the near-equipartition already noted by various authors \cite{Medvedev1999,SilvaApJ}.

\section{The initial field amplitude}\label{sec:Bi}
We now turn to the evaluation of the initial field amplitude. The idea is that the instability mechanism picks up a seed field from the spontaneous fluctuations of the system, and amplifies it. Starting with  Buneman, Salpeter and Sitenko \cite{Buneman,SalpeterFluctu,Sitenko}, various authors have been dealing with plasma fluctuations \cite{TajimaFluctu,LundFluctu,YoonFluctu,TreumannFluctu,TautzFluctu,SchlickeiserFluctu,Schlickeiser2012}.

Though initially focused on stable systems, fluctuation theory has been recently extended to weakly amplified modes \cite{TautzFluctu,SchlickeiserFluctu,Schlickeiser2012}. Note also that the ability of PIC simulations to correctly render them has been checked \cite{DieckmannFluctu}.

A first question to ask could be the following: should we consider the instability starts from the fluctuations of one single plasma, or from the fluctuations of the system formed by the two overlapping plasmas? In other words, should we consider the fluctuations of the system \textit{before} it turns unstable, or \textit{after}? We will now argue that we consider the fluctuations of the stable, isolated plasma shells, \textit{before} they interpenetrate.

Before they overlap, each plasma shell comes with its own fluctuations. As they start to overlap, the fluctuation fields for each plasma will adapt to each other. But on the very same time scale, the instability process begins. We thus consider the seeds for the instability are the ones which were already present in the system \textit{before} the plasmas started to overlap. Filamentation for example, needs unbalanced counter-streaming currents to start growing. As they approach each other, both plasma shells already display spontaneous fluctuations normal to the drift. When they start to interpenetrate, these fluctuations instantaneously provide the needed unbalanced currents to destabilize the whole system. Hence, their amplitude will be the amplitude they had \textit{before} they go unstable.

\subsection{Fluctuation power spectrum}
We are interested in the magnetic fluctuation spectra of a relativistically drifting, stable plasma. For the non-relativistic case, such calculation has been performed by Yoon \cite{YoonFluctu}. For the relativistic case, the amplitude of the spontaneous magnetic fluctuations for $k_\parallel=0$ can be deduced from  the linearized Vlasov-Maxwell system which yields the relation between the electric and current density fields \cite{Ichimaru},
\begin{equation}
\mathbf{j}(\mathbf{k},\omega) = \mathbf{Z}^{-1}(\mathbf{k},\omega)\cdot\mathbf{E}(\mathbf{k},\omega),
\end{equation}
where $\mathbf{Z}^{-1}$ is the tensor
\begin{equation}
(\mathbf{Z}^{-1})_{\alpha\beta} =\frac{i\omega}{4\pi} [\epsilon_{\alpha\beta} - \frac{k_\perp^2}{\omega^2}(1 - \delta_{\alpha y}\delta_{\beta y}) ] \, ,
\end{equation}
where it is assumed that the plasma drifts along the $x$-axis and the wave number $k_\perp$ is along the $y$-axis. There follows  \cite{Sitenko,Ichimaru},
\begin{equation}\label{fluctuations}
\mathbf{EE}^{\dagger}_{\omega,k_\perp}=
\mathbf{Z}_{\omega,k_\perp}\cdot\mathbf{jj}^{\dagger}_{\omega,k_\perp}\cdot\mathbf{Z}^{\dagger}_{\omega,k_\perp},
\end{equation}
where $\mathbf{EE}^{\dagger}_{\omega,k_\perp} $ is the fourier transform of the spontaneously emitted electric field fluctuation tensor and $\dagger$ designates the hermitian adjoint. Taking the $xx$ component of Eq. (\ref{fluctuations}) and dividing it by the square of the phase speed gives:
\begin{equation}
B^2_{\omega,k_\perp} = \frac{k_\perp^2}{\omega^2} [\mathbf{Z}_{\omega,k_\perp}\cdot\mathbf{jj^{\dagger}}_{\omega,k_\perp}\cdot\mathbf{Z}^{\dagger}_{\omega,k_\perp}]_{xx}.
\end{equation}

The dielectric tensor $\epsilon_{\alpha\beta}$ is given by \cite{Ichimaru},
\begin{eqnarray}
\epsilon_{\alpha\beta} &=& \delta_{\alpha\beta} \\
&+& \sum_s \frac{\omega_{ps}^2}{\omega^2}
\int d^3p\left[
\frac{p_\alpha}{\gamma(\mathbf{p})} \frac{\partial f_s^0}{\partial p_\beta}
+ v_\alpha \frac{p_\beta}{\gamma(\mathbf{p})}\frac{\mathbf{k}\cdot \partial f_s^0/\partial \mathbf{p}}{\omega-\mathbf{k}\cdot\mathbf{v}}
\right],\nonumber
\end{eqnarray}
where the sum runs along the 2 species involved, namely electrons and positrons. Both species are assumed to obey Maxwell-J\"uttner distribution functions \cite{Juttner1911,Wright1975},
\begin{eqnarray}
f_s^0(\mathbf{p}) &=& \frac{\mu}{4\pi\gamma_0^2K_2(\mu/\gamma_0)}\exp\left[-\mu\left(\gamma(\mathbf{p}) - \frac{\mathbf{v}_{0s}\cdot\mathbf{p}}{mc^2}\right)\right],\nonumber\\
\mu&=&\frac{mc^2}{k_B T},
\end{eqnarray}
where $K_2$ is the Bessel function of the second kind and $\mathbf{v}_{0s}$ the drift velocity of the s-th species. The evaluation of these quadratures by means of complex analysis techniques is extensive and quite involved, and will be reported in details in a separate paper \cite{RuyerGremillet}.

The $\omega$-integrated fluctuation energy density reads,
\begin{equation}\label{eq:Bi1}
\frac{B_{k_\perp}^2}{8\pi}=\frac{k_BT}{2} \frac{\frac{k_\perp^2c^2}{\omega_p^2} + \mu}
{\frac{k_\perp^2c^2}{\omega_p^2} + \mu/\gamma_0^2}.
\end{equation}
Since filamentation modes have $\omega=0$,  it is interesting to consider the spectrum density for $\omega=0$. In the regime $1\ll\gamma_0\ll \mu$ we consider here, and for $k_\perp=\omega_p/c\sqrt{\gamma_0}$, it reads \cite{RuyerGremillet},
\begin{eqnarray}\label{eq:Bi2}
\frac{B_{k_\perp,\omega}^2(\omega=0)}{8\pi} & \equiv & \frac{B_{k_\perp,0}^2}{8\pi}\\
&=&\frac{1}{\sqrt{32 \pi}}\frac{\gamma_0^3}{\sqrt{\mu}}\frac{mc^2}{\omega_p},\nonumber
\end{eqnarray}
Equation (\ref{eq:Bi1}) needs integration over a $\mathbf{k}$ domain to obtain the available power in the corresponding fluctuations, while result (\ref{eq:Bi2}) needs integration over a given $(\mathbf{k},\omega)$ domain.

One could assume the instability process is indeed a initial condition problem, so that it can discriminate the unstable $\mathbf{k}$'s, but not their frequency. In such case, Eq. (\ref{eq:Bi1}) should be used to derive the initial field amplitude. We here argue instead that the instability process can discriminate the fluctuation frequency, but up to a precision $\pm \delta$. In other words, $\omega=0$ is selected for growth, but this selection should be inaccurate to an order $\pm \delta$ because during the first growth period, the plasma cannot discriminate waves varying at $\omega=0\pm \delta$ from the ones at $\omega=0$. The two approaches will be later compared, and the ``$\omega=0$ selection'' will be found in slightly better agreement with the simulations.

\subsection{The $\mathbf{k}$-integration domain}
Whether we use Eqs. (\ref{eq:Bi1}) or (\ref{eq:Bi2}) for the available energy for growth, we thus need to integrate over a $\mathbf{k}$ domain. As indicated on Fig. \ref{fig:taux}, wave-vectors selected for growth form a narrow band around the normal axis, extending to infinity and of width
\begin{equation}\label{eq:kmax}
Z_\parallel\sim \sqrt{\frac{2}{\gamma_0}} \Longrightarrow k_{\parallel,max}\sim \sqrt{\frac{2}{\gamma_0}}\frac{\omega_p}{c},
\end{equation}
in the parallel direction (we set here $v_0\sim c$).

Regarding the integration domain in the normal direction, it has already been mentioned that the fastest growing mode has been numerically found for $k_\perp\sim \omega_p/c\sqrt{\gamma_0}$. We shall thus integrate Eqs. (\ref{eq:Bi1},\ref{eq:Bi2}) from $k_{\perp, min}$ to $k_{\perp, max}$ with,
\begin{eqnarray}\label{eq:kperps}
    k_{\perp, min}=\frac{1}{2}\frac{\omega_p}{c\sqrt{\gamma_0}},\nonumber\\
    k_{\perp, max}=2\frac{\omega_p}{c\sqrt{\gamma_0}},
\end{eqnarray}
where the factors 1/2 and 2 have been arbitrarily chosen to bracket $k_m=\omega_p/c\sqrt{\gamma_0}$. Note that the end result is almost independent of these constants because of the logarithm function in Eq. (\ref{eq:Bf}).

\subsection{Saturation time from $\omega$-integrated fluctuations}
The $\omega$-integrated energy density (\ref{eq:Bi1}) is  now integrated in the following way,
\begin{equation}\label{eq:quadraBi1}
    \frac{B_i^2}{8\pi}=\int_{k_{\perp, min}}^{k_{\perp, max}} 2\pi k_\perp dk_\perp \int_{-k_{\parallel,max}}^{k_{\parallel,max}}dk_\parallel
    \frac{B_{k_\perp}^2}{8\pi}.
\end{equation}
Clearly, it is an \emph{averaged} initial amplitude over the modes likely to grow the most. The result reads,
\begin{equation}
    \frac{B_i^2}{8\pi}= \pi \sqrt{\frac{2}{\gamma_0}}\left(\frac{15}{4\gamma_0}+\mu  \ln\left[\frac{1+4 \gamma_0/\mu }{1+\gamma_0/4\mu }\right]\right)\left(\frac{\omega_p}{c}\right)^3 k_BT.
\end{equation}
Inserting this result in Eq. (\ref{eq:Bf}) for the saturation time, we find
\begin{eqnarray}\label{eq:tauOK1}
\tau_s\omega_p&=&\frac{\sqrt{\gamma_0}}{2 \sqrt{2}}\ln\left[
\frac{n(c/\omega_p)^3}{\sqrt{2} \pi } \frac{\mu\gamma_0^{3/2}}{\frac{15 }{4 \gamma_0 }+
\mu  \ln\left[\frac{1+4 \gamma_0/\mu }{1+\gamma_0/4 \mu }\right]}
\right] \\
&\sim& \frac{\sqrt{\gamma_0}}{2 \sqrt{2}}\ln\left[
\frac{2 \sqrt{2}}{15 \pi } n\left(\frac{c}{\omega_p}\right)^3 \sqrt{\gamma_0 } \mu\right],~~\mathrm{for}~~ \gamma_0\ll \mu,\nonumber
\end{eqnarray}
where $n$ is the plasma density.

\subsection{Saturation time from fluctuations near $\omega=0$}
While the $\mathbf{k}$-integration domain remains unchanged, an $\omega$-integration domain is now needed. As stated earlier, physical reasoning would suggest an integration over $[-\delta,\delta]$, because the instability mechanism cannot discriminate fluctuations with $\omega=0$ from the ones with $-\delta<\omega<\delta$ during the first growth period or so. Once a given fluctuations has been significantly amplified, i.e, has grown during $\sim\delta^{-1}$, it will keep on growing.

But the integration domain is eventually much smaller than $[-\delta,\delta]$ because the density $B_{k_\perp,\omega}$ is extremely peaked around $\omega=0$ with,
\begin{equation}
B_{k_\perp,\omega} = B_{k_\perp,0}\left(1-\frac{\omega^2}{\delta\omega^2}+o(\omega^2)\right).
\end{equation}
At $k_\perp=\omega_p/c\sqrt{\gamma_0}$, the peak width $\delta\omega$ is given by \cite{RuyerGremillet}
\begin{equation}
\delta\omega \sim \frac{\omega_p}{\gamma_0\sqrt{6\mu}},
\end{equation}
which turns out thinner than the growth rate $\delta=\omega_p\sqrt{2/\gamma_0}$, especially for $\mu\gg 1$. The energy density (\ref{eq:Bi2}) is therefore integrated over $[-\delta\omega,\delta\omega]$, and the initial field amplitude $B_i$ from the energy available for growth computed as,
\begin{equation}\label{eq:quadraBi}
    \frac{B_i^2}{8\pi}=\int_{k_{\perp,min}}^{k_{\perp,max}} 2\pi k_\perp dk_\perp \int_{-k_{\parallel,max}}^{k_{\parallel,max}}dk_\parallel\int_{-\delta\omega}^{\delta\omega}d\omega
    \frac{B_{k_\perp,0}^2}{8\pi}.
\end{equation}

Given the narrowness of the $(\mathbf{k},\omega)$ integration domain, we set $B_{k,\omega}\sim B_{k_\perp,0}$  as given by Eq. (\ref{eq:Bi2}) in the integrand. A little algebra gives
\begin{equation}\label{eq:BiOK}
 \frac{B_i^2}{8\pi}=\frac{15 \sqrt{\pi/6}}{4}\frac{\sqrt{\gamma_0}}{\mu} \left(\frac{\omega_p}{c}\right)^3 m c^2 ,
\end{equation}
and the saturation time can be cast under the form
\begin{equation}\label{eq:tauOK2}
\tau_s\omega_p=\frac{\sqrt{\gamma_0}}{2 \sqrt{2}}
\ln\left[
\frac{4}{15}\sqrt{\frac{6}{\pi}}n\left(\frac{c}{\omega_p}\right)^3 \sqrt{\gamma_0}  \mu \right].
\end{equation}

Finally, it is to be reminded that our theory has been implemented for a 3D geometry, whereas simulations are 2D. The corresponding 2D saturation time is derived in Appendix \ref{ap:2} and reads,
\begin{equation}\label{eq:tau2D}
\tau_s\omega_p=\frac{\sqrt{\gamma_0}}{2 \sqrt{2}}
\ln\left[
4\times 10^2 \sqrt{\frac{\pi}{3}} \frac{\mu}{\gamma_0} N
\right],
\end{equation}
where $N$ is the number of macro-particles per cell.

\section{Comparison with Simulations} \label{sec:sim}
In order to test the theory in the early stage of shock formation, ab-initio Particle-In-Cell simulations have been performed \cite{Osiris1,Osiris2}. The shock is launched with the piston-wall method, where two counter-propagating symmetric plasma beams are produced by injecting one beam by a cathode from one side of the simulation box, and being reflected at the opposite wall. Here we simulate pair plasmas with low temperature parameter $\mu = mc^2 / k_B T = 10^6\gamma_0$, and Lorentz factor \(\gamma_0 \in [25, 10^4]\), so that \(1 \ll \gamma_0 \ll \mu\) is fulfilled. The particles are injected along the \(x\) axis with a temporal resolution \(\Delta t = 0.025 \sqrt{\gamma_0} /\omega_p\) and the size of a cell being \(\Delta x = 0.05 \sqrt{\gamma_0} c/ \omega_p\), using quadratic interpolation and 8 particles per cell. Note that this results in a grid size much larger than the Debye length, which could trigger the grid instability \cite{bird85}. However, for the combination of grid sizes and number of particles per cell we have used, this instability has a much longer time scale than the typical times analyzed in our simulations. Moreover, in our simulations we use higher order particle shapes and current smoothing (a 5 pass binomial smooth is used).
This improves significantly the energy conservation properties of the algorithm and slows down even further numerical heating \cite{Osiris2}. The present spatial resolution has been tested in many shock simulations confirming its applicability. Convergence tests have been performed with smaller cell sizes, and no significant deviations was found.
Physically, any numerical heating arising from the grid instability is irrelevant to the dynamic of the problem because the free energy in the system converted to thermal energy is much larger than the energy associated with numerical heating.

We ran test simulations with up to 800 particles per cell and observed only small deviations to the results reported here (see Fig. \ref{fig:form800}). Detailed convergence tests were also performed. The two-dimensional box with \( L_x = 125 \sqrt{\gamma_0} c/ \omega_p\) and \( L_y = 5 \sqrt{\gamma_0} c/ \omega_p\) has absorbing boundaries for the particles along \(x\) and is periodic along \(y\). For the fields, conducting boundaries are used at the perfectly reflecting wall and open boundary conditions at the cathode.

Since we are interested in the early stage of shock formation, the question arises as to whether the piston-wall method is appropriate. We first simulate the periodic system of counter-streaming beams, corresponding to the model of unstable fluctuations that is the basis of our theoretical approach. In this case, no shock is formed and we are able to identify the growth rate and saturation time of the magnetic field energy.
We compare the periodic system with the piston-wall setup and, furthermore, with the full shock formation process, where in \(x\) direction absorbing boundaries have been used for the particles and conducting boundaries for the fields. In the latter case, two symmetric shocks are propagating outwards and this allows us to identify non-physical fields at the reflecting wall in the piston-wall setup.

\begin{figure}
\begin{center}
	\includegraphics[width=0.45\textwidth]{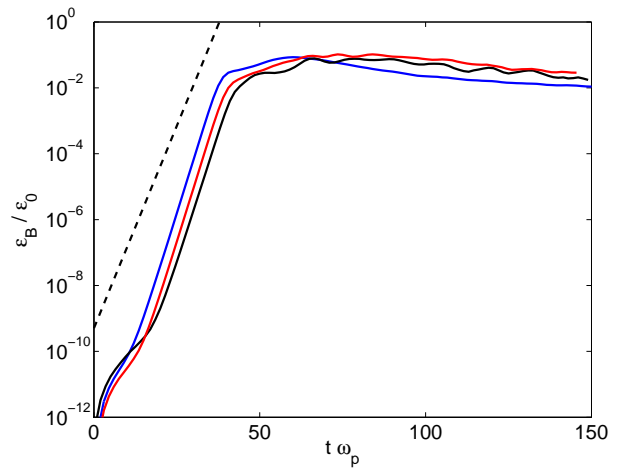}
\end{center}
\caption{Magnetic field energy evolution for different simulation setups and \(\gamma_0 = 25\). \emph{Black}: piston-wall method, \emph{Red}: full shock picture, \emph{Blue}: periodic system of counter-streaming beams, \emph{Black dashed}: theoretical growth rate. A detailed description of the models is given in the text.} \label{fig:setups}
\end{figure}

\begin{figure}
\begin{center}
\includegraphics[width=0.45\textwidth]{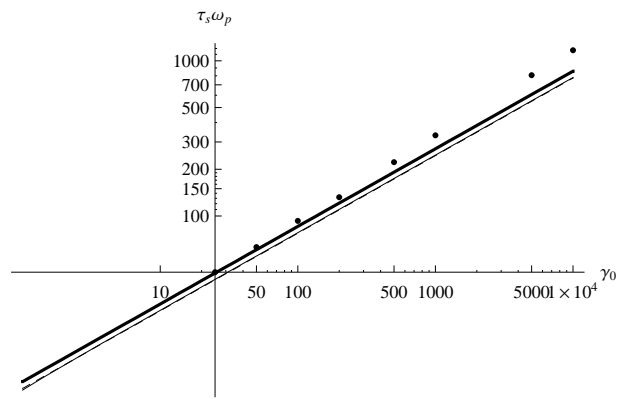}
\end{center}
\caption{Saturation time $\tau_s\omega_p$ from the PIC simulations, \emph{circles}, from the fluctuations near $\omega=0$ Eq. (\ref{eq:tauOK2}), \emph{bold line}, from the $\omega$-integrated fluctuations Eq. (\ref{eq:tauOK1}), \emph{thin line}, and from the 2D formula (\ref{eq:tau2D}), \emph{thin dashed line}. The 3D $\omega$-integrated and the 2D theories give almost the same result.
} \label{fig:compa}
\end{figure}

Fig. \ref{fig:setups} shows the evolution of the magnetic field energy \(\epsilon_B\) normalized by the kinetic energy in the box at time zero \(\epsilon_0\) for the three different setups for \(\gamma_0= 25\). The comparison shows that the growth rate and saturation level of the field is independent of the setup. The theory of the periodic system applies to the non-periodic system as well, where the overlapping beams go unstable, and the fields at the reflecting wall do not seem to affect this process. There is only a small deviation in the initial fluctuation level, which for \(\gamma_0= 25\) leads to a shift of the saturation time $\sim \omega_p^{-1}$ between the different setups. On the interesting time scales for the saturation time (see Fig.\ \ref{fig:compa}) this deviation is negligible, so that we conduct the simulations with the piston-wall setup in order to save simulation time.

Theoretical results are now bridged setting
\begin{equation}
	n \left( \frac{c}{\omega_p} \right)^3 = \frac{8}{0.05^3 \, \gamma_0^{3/2}}.
\end{equation}
Figure \ref{fig:compa} compares the saturation time measured in the simulations with Eq. (\ref{eq:tauOK2}) accounting for fluctuations near \(\omega = 0\), Eq. (\ref{eq:tauOK1}) accounting for \(\omega\)-integrated fluctuations and the 2D formula (\ref{eq:tau2D}). As expected, considering only the fluctuations around \(\omega = 0\) yields a larger saturation time, arising from a lower initial noise amplitude. The slight underestimation of the simulation results can be attributed to at least two factors. On the one hand, Eq. (\ref{eq:tauOK1}) necessarily remains a lower limit, as the integration domain only brackets the mode selected for growth. On the other hand, it is difficult to model the level of fluctuations in the simulations realistically, since it is dependent and sensitive on the choice of numerical parameters of the simulations.

\section{Conclusion}
In this paper, we have analyzed in detail the first part of a collisionless shock formation process. We chose the simplest possible setup for analytical calculations. We have thus run relativistic PIC simulations of two interpenetrating cold pair plasmas shells, the initial Lorentz factor being the only varying parameter. In the present collisionless conditions, these shells could simply pass through each other. But because the overlapping region is unstable,  turbulence is triggered which eventually leads to the shock formation.

The shock formation time has been determined from the expression of the dominant growth rate, the field at saturation and the seed field amplified by the instability.  The dominant growth-rate could be determined from the theory derived for an homogeneous, infinite system, in spite of the limited extension of the overlapping region. The agreement between the simulations and our simple model is due to the fact that although finite, the center of the unstable region, where modes start growing, satisfies the homogeneity criterion.

The field at saturation is correctly given by any of the 3 existing saturation criteria, as the wave number of the dominant unstable mode precisely adapts for these criteria to converge (up to a numerical constant). The reason why the system ``chooses'' to amplify preferably this $k_\perp$, in spite of the absence of a peak in the growth-rate curve $\delta(k_\perp)$ could be the topic of future works.

Finally, the seed field $B_i$ which the instability picks up for amplification is computed from the amplitude of the spontaneous fluctuations of one single relativistically streaming shell. On the one hand, this density is integrated over the $\mathbf{k}$ domain likely to grow. On the other hand, we have tested the $B_i$ value obtained assuming the instability mechanism purely acts as an initial value process, unable therefore to discriminate the frequency of the noise, and the $B_i$ value obtained assuming the instability selects for amplification those fluctuations with $\omega=0$. By computing the saturation time given by both options, we find the second one fits slightly better the simulations.

The reasoning used to time this first phase can in principle be adapted to any settings. By the end of the linear growth phase, the density of the overlapping region is still about twice the upstream density, as evidenced on Fig. \ref{fig:form}. Indeed, because the linear regime requires small perturbations, it is necessarily over by the time density perturbations reach $\delta n\sim 0.1-1$. For the present system, the density jump around the shock soon to be formed is around 3. This implies other processes have to pick-up the system from the saturation time up to the shock formation. We plan to dedicate future works to these mechanisms.

\appendix

\section{Oblique to filamentation transition}\label{ap:1}
It can be seen from Fig. \ref{fig:taux} that the growth rate at large $Z_\perp$ reaches a limit $\delta_{Z_\perp,\infty}$ which is function of $Z_\parallel$. For $\gamma_0=1.1$, $\delta_{Z_\perp,\infty}(Z_\parallel)$ reaches an extremum for $Z_\parallel\neq 0$, which corresponds to a spectrum governed by oblique modes. Then, for $\gamma_0=10$, the extremum is reached at $Z_\parallel =0$, and filamentation dominates. The first derivative $\partial \delta_{Z_\perp,\infty}/\partial Z_\parallel$ always vanishes for $Z_\parallel =0$. The transition from one regime to the other occurs then when the second derivative vanishes at $Z_\parallel =0$.

The asymptotic dispersion equation for $Z_\perp=\infty$ can be determined and reads,
\begin{equation}
    4(1- \gamma_0^2)-2 (x^2+Z_\parallel^2) \gamma_0+(x^2-Z_\parallel^2)^2 \gamma_0^4=0.
\end{equation}
This equation can be solved, and the growth rate for $Z_\perp=\infty$ is,
\begin{equation}
\delta_{Z_\perp,\infty}^2=Z_\parallel^2+\frac{1-\sqrt{1+4 \gamma_0^3  (Z_\parallel^2+\beta ^2 \gamma_0)}}{\gamma_0^3}.
\end{equation}
Deriving twice the expression above with respect to $Z_\parallel$ gives the Lorentz factor for the transition from the oblique to the filamentation regime,
\begin{eqnarray}
\gamma_0&=&\sqrt{\frac{3}{2}}\sim 1.22,\nonumber\\
\beta_0&=&\frac{1}{\sqrt{3}}\sim 0.57~.\nonumber
\end{eqnarray}

\section{Application to a 2D PIC plasma}\label{ap:2}
Care must be taken when using the formula (\ref{eq:Bi2}) for a 2D PIC-modeled plasma. The plasma is then composed of macro-particles with charge and mass equal, respectively,
to $Q_p = W_p q$ and $M_p = W_p m$, where $q$ and $m$ denote the real particles' charge and mass, and $W_p$ is the statistical weight. In a 3D plasma, $W_p$ is a
dimensionless quantity,  whereas it is a lineic density in 2D. For the numerical plasma to behave collectively as its physical counterpart, the plasma frequencies of the two
systems must be equal, which implies
\begin{equation} \label{eq:Wp}
  W_p = \frac{m\omega_p^2}{4\pi q^2}\frac{\Delta x \Delta y}{N} \, ,
\end{equation}
where $N$ is the number of macro-particles per cell and $\Delta x = \Delta y$ is the cell size.

In a 2D geometry, the fluctuation field is then given by
\begin{align}\label{eq:Bi_pic}
  B_i^2 &\sim\int_{k_\perp, min}^{k_\perp, max} dk_\perp \int_{-k_{\parallel, max}}^{k_{\parallel, max}} dk_\parallel \int_{-\delta \omega}^{\delta \omega} B_{k,\omega=0}^2 \nonumber \\
  &\sim \frac{12}{\sqrt{3}} \frac{\gamma_0}{\mu}\left( \frac{\omega_p}{c} \right)^2 W_p m c^2 \, .
\end{align}
Note that the normalized inverse temperature $\mu$ is also an invariant.  Substitution of Eq. \eqref{eq:Wp} and  $\Delta x = 0.05 \sqrt{\gamma_0}c/\omega_p$ into
\eqref{eq:Bi_pic} readily yields
\begin{equation}
  B_i^2 = \frac{2.5\times 10^{-3}}{N} \sqrt{\frac{3}{\pi}} \frac{\gamma_0^2}{\mu} \left(\frac{mc\omega_p}{q}\right)^2 \, .
\end{equation}
There follows the ratio
\begin{equation}
  \frac{B_f^2}{B_i^2} =  4\times 10^2 \sqrt{\frac{\pi}{3}} \frac{\mu}{\gamma_0} N
\end{equation}
and the saturation time given by Eq. (\ref{eq:tau2D}).

\acknowledgments
This work was supported by projects ENE2009-09276 of the Spanish Ministerio de Educacio´n y Ciencia, the European Research Council (ERC-2010-AdG Grant 267841) and FCT (Portugal) grants PTDC/FIS/111720/2009 and SFRH/BD/38952/2007. Thanks are due to Lorenzo Sironi for useful discussions. A.B. wishes to thank the Harvard-Smithsonian Center for Astrophysics for hosting him.

\bibliography{Bib}

\end{document}